\documentclass[prb]{revtex4}   

\usepackage{amsmath}    
\usepackage{graphicx}   

\begin{document}

\title{Astrophysics datamining in the classroom: Exploring real data with new software tools and robotic telescopes}
 
\author{Rosa Doran}
 \affiliation{NUCLIO - N\'ucleo Interactivo de Astronomia, Largo dos Top\'azios, 48 $3^o$ Frente, 2785 - 817 S\~ao Domingos de Rana, Portugal}
 \email{rosa.doran@nuclio.pt}   

\author{Anne-Laure Melchior}
\affiliation{Universit\'e de Paris VI Pierre et Marie Curie, 4 Place Jussieu, 75252 - Cedex 05 Paris, France}

\author{Thomas Boudier}
\affiliation{Universit\'e de Paris VI Pierre et Marie Curie, 4 Place Jussieu, 75252 - Cedex 05 Paris, France}

\author{Pac\^ome Delva}
\affiliation{Universit\'e de Paris VI Pierre et Marie Curie, 4 Place Jussieu, 75252 - Cedex 05 Paris, France}

\author{Roger Ferlet}
\affiliation{Institut d'Astrophysique de Paris, UMR7095 CNRS / UPMC, 98bis Bd Arago, FR 75014 Paris, France}

\author{Maria L.T. de Almeida}  
\affiliation{N\'ucleo Interactivo de Astronomia, Largo dos Top\'azios, 48 $3^o$ Frente, 2785 - 817 S\~ao Domingos de Rana, Portugal}

\author{Domingos Barbosa}
 \affiliation{Grupo de Radio Astronomia, Instituto de Telecomunica\c c\~oes, U. Aveiro, 3810-193 Aveiro, Portugal}

\author{Edward Gomez}
\affiliation{Las Cumbres Observatory Global Telescope Network, Inc., 6740 Cortona Drive, Suite 102, Goleta, CA 93117,USA}

\author{Carl Pennypacker}
\affiliation{Hands On Universe Division: Physics/SSL, Lawrence Berkeley National Lab, 1 Cyclotron Road, Mail Stop 50-232, Berkeley, CA 94720, USA}  
\author{Paul Roche}
\affiliation{Faulkes Telescope Project, Astronomy Centre, HESAS, University of Glamorgan, Glyntaf, Wales, UK. CF37 4BD, United Kingdom}
\author{Sarah Roberts}
\affiliation{Faulkes Telescope Project, Astronomy Centre, HESAS, University of Glamorgan, Glyntaf, Wales, UK. CF37 4BD, United Kingdom}

\date{\today}

\begin{abstract}
Within the efforts to bring frontline interactive astrophysics and astronomy to the classroom, the Hands on Universe (HOU) developed a set of exercises and platform using real data obtained by some of the most advanced ground and space observatories. The backbone of this endeavour is a new free software Web tool - {\bf S}uch {\bf a} {\bf L}ovely {\bf S}oftware for {\bf A}stronomy based on Image{\bf J} (SalsaJ). It is student-friendly and developed specifically for the HOU project and targets middle and high schools. It allows students to display, analyze, and explore professionally obtained astronomical images, while learning concepts on gravitational dynamics, kinematics, nuclear fusion, electromagnetism. The continuous evolving set of exercises and tutorials is being completed with real (professionally obtained) data to download and detailed tutorials. The flexibility of the SalsaJ platform tool enables students and teachers to extend the exercises with their own observations through the use of robotic telescopes. The software developed for the HOU program has been designed to be a multi-platform, multi-lingual experience for image manipulation and analysis in the classroom. Its design enables easy implementation of new facilities (extensions and plugins), minimal in-situ maintenance and flexibility for exercise plugin. Here, we describe some of the most advanced exercises about astrophysics in the classroom, addressing particular examples on gravitational dynamics, concepts currently introduced in most sciences curricula in middle and high schools.
\end{abstract}

\maketitle

\section{Introduction}

In recent years, many studies have highlighted a decline in students' and young people's interest in science. Revitalizing the teaching of sciences and in particular physics through astronomy examples with hands-on activities and the learning of scientific methods could make scientific subjects more attractive in university and school education.  

Connections to learning communities and in particular classroom engagement need to be complemented by a "plug and play", hands-on approach with simple, yet high-performance instruments coupled to web interfaces which are capable of user-friendly remote exploration and tied into Virtual Observatory\cite{vo1,vo2,vo3,vo4} databases. As most of our activities are based on data, it is essential to provide analysis tools well adapted to the pupil-classroom framework. Access to Virtual Observatories data sets is now possible for anyone, since they offer a standardization and cross data centres interoperability of the data formats, access technologies and professional software tools for the purpose of efficient and unrestricted access to astronomical information and knowledge. However, the need for a common, easy to use and flexible software platform to manipulate and datamine real astrophysical data from the great observatories in ground and space led to the development of SalsaJ\cite{euhou, salsaj}, an open-source, web platform software itself based on the Image Processing and Analysis in Java (ImageJ)\cite{imageJ,imj1,imj2} software. ImageJ itself was designed with an open architecture that provides extensibility and modularity via custom made Java plugins. 

The pioneering experience of the Berkeley initiated Hands-On Universe (HOU)\cite{hou}, an educational program founded by Carl Pennypacker at Berkeley, became globalized through the Global Hands-On Universe (GHOU)\cite{ghou} movement by joining efforts from all around the world, allowing students and teachers to share their work and experiences. HOU started to promote the exploitation of the wealth of data being produced in astronomy and introduced mainstream astrophysical science problems (data and tools) into learning activities. Examples are the Supernovae anaylsis exercises, a subject awarded with a Physics Nobel prize. Among others HOU had the collaboration of the 2006 and 2011 Physics Nobel Prize Winners George Smoot and Saul Perlmutter. 

HOU efforts led to the development of SalsaJ, enabling students to investigate the Universe while applying tools and concepts from science, math, and technology.  In particular, SalsaJ was developed as a modular platform, allowing analysis of professionally obtained  astrophysical data with the aid of user-friendly image processing software via the Internet and can accommodate new user written plugins developed meanwhile by the European part of HOU (EU-HOU)\cite{euhou}. These two facts (real data and plug-in modularity) constitute the major differences from the plethora of software platforms for astronomy exploration making it a unique and suitable tool for schools. 

Most of the available astronomical software platforms will fall on the planetarium type like Stellarium\cite{stellarium} or Celestia\cite{stellarium} to name a few, for simulation of sky or Solar system exploration but without the capabilities to actually introduce physics concepts and manipulate real data. Also, other approaches will include more traditional astronomical observations activities, relying on the excellent capabilities of amateurs astronomy and even game strategies, enabling teachers to engage students in the exploration of certain topics and evaluate their knowledge. An example of such is a recent project, Universe Quest (UQ, now freely available for educators). The main goal of UQ is to awaken the interest of science in girls. In Portugal, a pilot effort was initiated involving 6 schools and students from the 7th to the 12th grade, in classroom environment. The tool can be used to promote Inquiry-based science education (IBSE) while using modern 3d assets\cite{universequest}. Although this safe environment is very rich in content and capabilities, it still lacks the capability to actually perform a scientific activity in a close to professional, but adapted classroom environment.

While the SalsaJ platform targets middle and high school classrooms (up to K-12 level in the USA) the level and accessibility of many exercises, and the richness of real space and ground observatory data available on SalsaJ database, makes it a versatile tool for college level introductory lectures on physics and astronomy. In fact, any teachers or scholars may develop their own Java plugin according to the desired complexity of an exercise. Using this tool students can learn about CCD cameras, telescopes and their applications and other technical aspects related to imaging. Besides its unique capabilities, SalsaJ backbone is currently maintained by the University of Paris VI Pierre et Marie Curie and constitutes a multi-lingual platform with translations to many languages: English, French, Spanish, Italian, Polish, Greek, Portuguese, Swedish, Northern Sami, Arabic, Chinese.  

The flexibility and modularity of SalsaJ allow many exercises and tutorials covering topics in physics and mathematics to be addressed through the analysis of real astrophysical data. Examples range from observing and analyzing extra-solar planet parameters to measuring the mass of a black hole in the center of our galaxy while applying Kepler’s Laws. Teachers can give these exercises and data to their students to illustrate and enliven physics concepts. On a scientific point, students can learn mathematics and physics while understanding the measurement of the brightness of a star and the flux laws, nuclear fusion, chemical elements through spectroscopy, the shape of a galaxy or trace the trajectory of an asteroid (kinematics, gravitational dynamics). 

SalsaJ relies on real astronomical data and on the expertise of scientists to produce material, which has been previously tested and is ready to use in schools. It also provides access to remote observing tools in radio wavelengths and in the optical. These research-quality tools, developed for education, enable the pupils to acquire their own data (images or spectra) in the classroom through a web interface. In parallel, a webcam system, developed by the Polish partner (CFT, Warsaw, Poland) and his team for astronomical observations (Koralewski et al. 2003, Kalicki et al. 2004), enables local acquisition of data. SalsaJ is now a backbone  project to integrate tools in an online teaching and learning platform, STOCHASMOS\cite{stochasmos1,stochasmos2}. This will improve a readily integration into the design of learning environments for reflective inquiry and guide inquiry-science teaching in classrooms.

\begin{figure}
\centering
\includegraphics[width=3.5in]{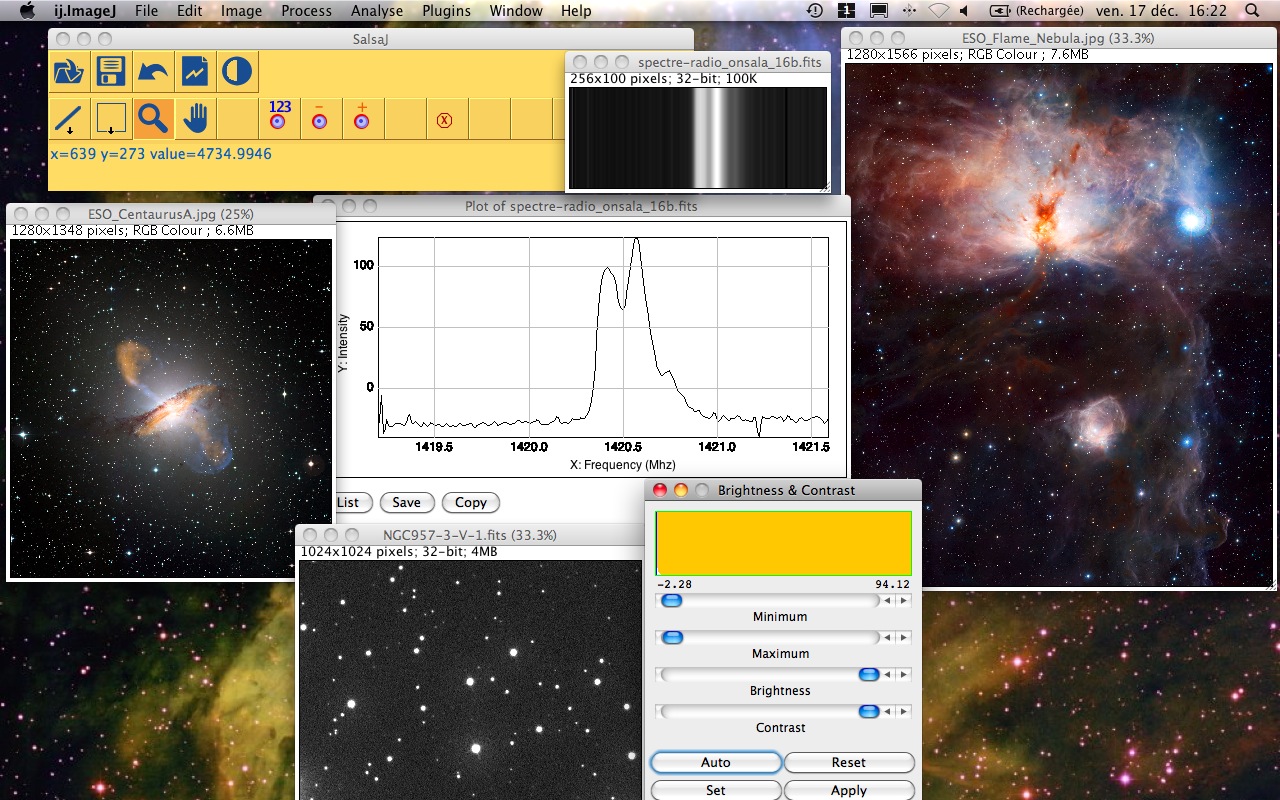}
\includegraphics[width=3.5in]{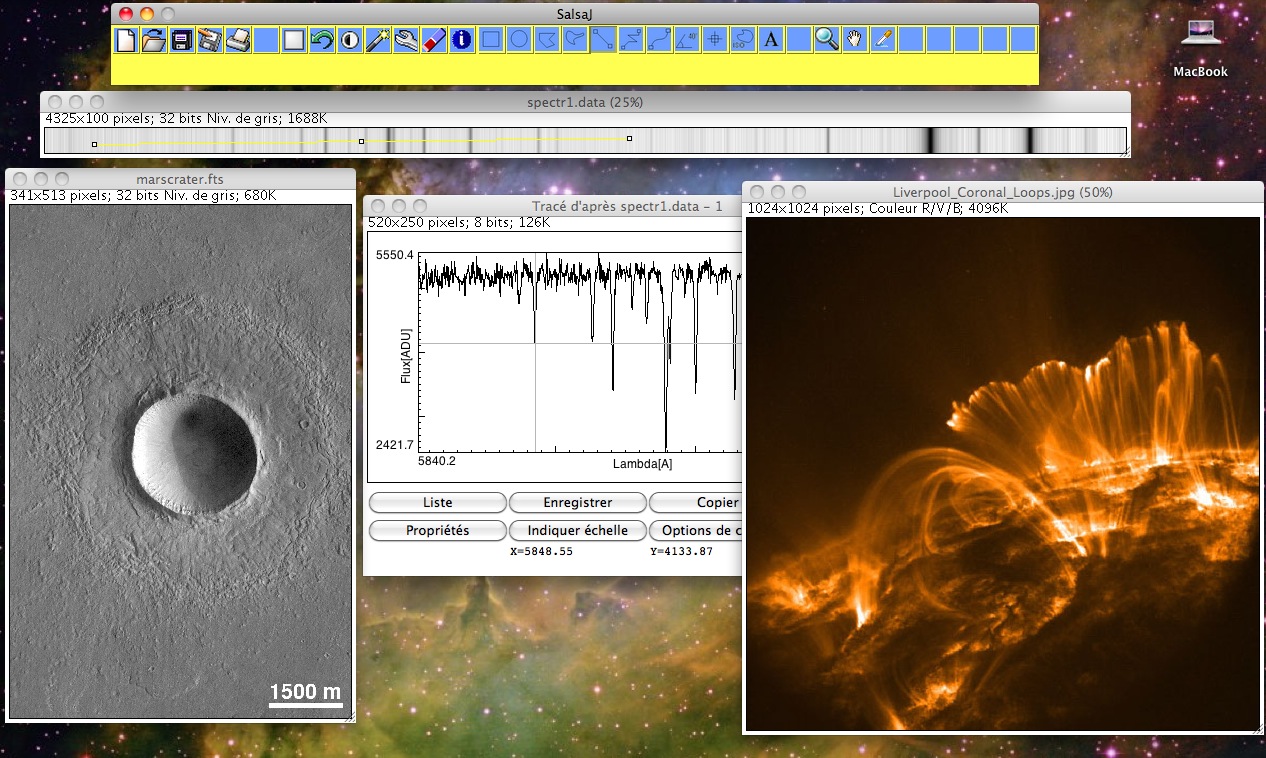}
\caption{Print screen examples of exercise sessions with SalsaJ. Left: Solar system exploitation - finding crater sizes; Right: SalsaJ galaxy images and spectra manipulation - one is Centaurus A: This galaxy is a most interesting object for the present attempts to understand active galaxies. It is being investigated by means of observations in all spectral regions, from radio via infrared and optical wavelengths to X- and gamma-rays, thus constituting an excellent example for educational purposes.}
\label{fig1}
\end{figure}

\section{Material and Methods}
SalsaJ is based on the ImageJ\cite{imageJ} codebase, the public domain Java Image processing software developed by the US National Institute of Health for easy manipulation of medical data by students, but with custom made tool plugins with new functionalities added. SalsaJ inherits ImageJ world's fastest pure Java image processing program as it can filter a 2048x2048 image in less than 0.1 seconds. That's about 40 million pixels per second! The interface has been modified in order to be more pupil-friendly and target astronomy data. It includes colour icons for frequently used operations such as files opening or brightness/contrast adjustments. The menus have been modified, some sophisticated native tools have been removed while simple applications for astronomy have been added enabling some otherwise complex operations on image processing to proceed easily. SalsaJ, like ImageJ, will install and run on most major Operating Systems (Windows, Linux, Mac OS), with functionalities to process, display, edit, analyze, process, save and print 8-bit up to 32-bit images. Besides the most widely used image formats it can read DICOM, FITS, "raw" professional science data formats and supports "stacks" of images (a series of images that share a single window). ImageJ based tools provide area and pixel value statistics of user-defined selections; measurement of distances and angles, and support standard image processing functions. The SalsaJ concept exercises will fall most of the time into these categories:
\begin{itemize}
\item {\bf Aperture photometry}: In astronomy, stellar objects are convoluted by a transfer function (Earth's atmosphere and telescope optics) called the Point Spread Function (PSF) that can be well approximated by a Gaussian function. It is necessary to be able to measure the intensity of such objects. The most robust procedure used in astronomy is aperture photometry. The stellar intensity and the sky background are integrated in a disk centered on the star, while the sky background is estimated in a ring beyond the stellar radius. An automatic procedure is proposed in SalsaJ, as well as an access to the main parameters for a more advanced level.
\item {\bf Spectroscopy} : Along with images, astronomical data in the form of 1D spectroscopic data are available. A Gaussian fit for the integration of spectral lines has been added. These facilities will enable the pupils to study the Doppler effect with the detection of extra-solar planets, the rotation of galaxies or the study of the atomic hydrogen in the Milky Way. 
\item {\bf Data acquisition} : A Webcam plugin enables acquisition of data from modified Webcams directly through SalsaJ. The modified Webcams allow short and long exposures mode of integration of sky objects.
\end{itemize}
Furthermore, it is easy to analyze astronomical CCD obtained images: pixel size is defined by the physical size of pixel (e.g.$15\mu$m) and angular size in sky (e.g. 0.3 arcsec), ie depends on optical path followed by photons provided with image. 

These characteristics can be used to study scaled images to search sizes of certain features of astronomical objects. Of particular visual impact, planetary images are ideal to look for dimensions and size of craters.
This possibility is behind a modern trend of data mining where citizens are invited to help scientists analyze hundreds of survey images in search for new findings. The beauty of this system is that anyone possessing basic knowledge on how to analyze these images can benefit from this possibility. With the support of professional scientists classrooms can obtain new scientific results while learning curriculum content and acquiring important key skills.  We list here just a few of the most successful examples, leaving other detailed material in the references for further exploitation.

In parallel, a new class of robotic telescopes became available for online remote operation by users in their regular classroom. The Faulkes Telescopes, a pair of telescopes with 2-meter mirrors are open to the school community under special circumstances and complement data exercises available on SalsaJ databases. 

\section{Some Astrophysical exercises with real data}
\subsection{Physics of impact: craters from meteorites} The presence of craters is the most common land form feature on the surface of all telluric planets and icy satellites of the Solar System.  They constitute a signature of the presence of meteorites and its collision with planets. These bodies carrying organic molecules have crossed the Solar System and may have played an important role in the emergence of life conditions. An example may start with an image of Mars Crater PIA02084. This image was obtained in July 1999 with the Mars Orbiter camera on board of Mars Global Surveyor. This crater has been formed by the impact and explosion of a meteorite on the North of Elysium Planitia. 
\begin{figure}
\centering
\includegraphics[width=6.0in]{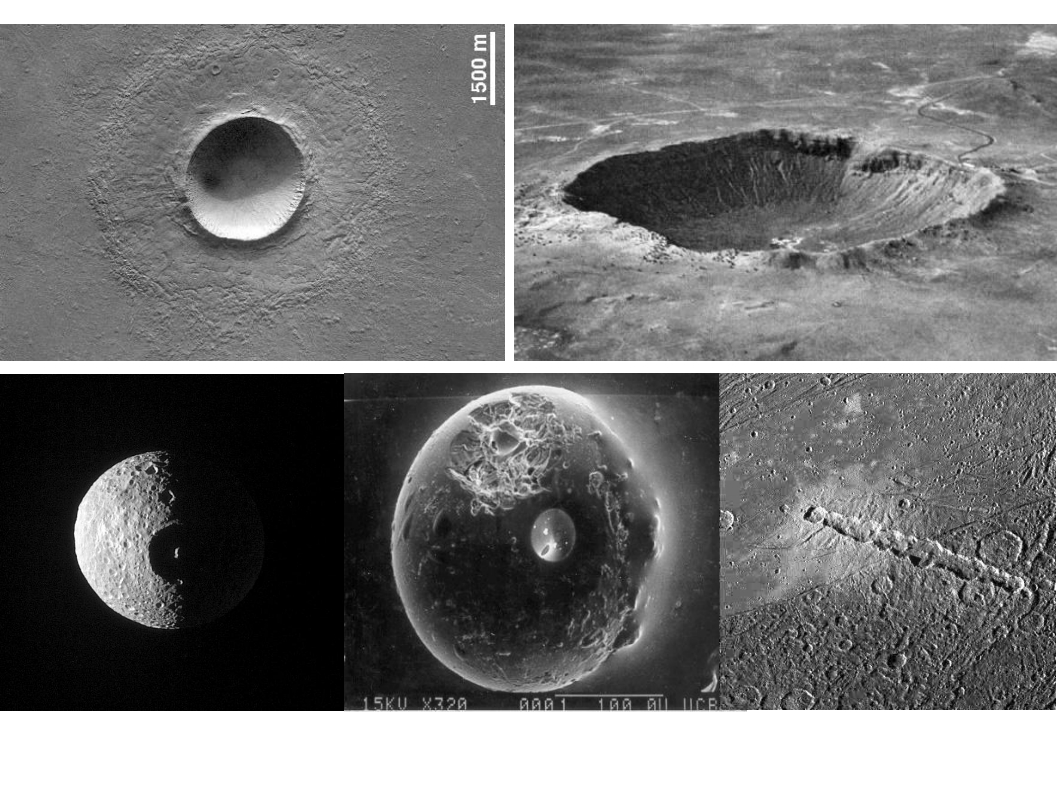}
\caption{From left to right and top to bottom:  Mars PIA02084 crater (@ NASA/JPL/Malin Space Science Systems); Barringer Meteor Crater, in Arizona (@ US Geological Survey); Mimas, a Saturn moon, from Cassini spacecraft in 2005 (@ NASA) and its Herschel crater; microscopic spherule imaged from an impact crater on the Moon brought back by the Apollo 11 (@ NASA); Ganymede surface as seen from Galileo spacecraft (@ NASA).}
\label{fig}
\end{figure}
From the image, students measure the size of the pixel in meters and derive the size of the crater. A comparison is then made with Earth craters like the Barringer Meteor Crater, in Arizona. This crater is thought to have been formed 25000-50000 years ago and its impactor, probably constituted of iron, has a diameter of the order 30-100 m, and a mass of 60000 tons. An estimation of the energy of impact can be found trough the calculation of the impactor kinetic energy ($E=\frac{1}{2}m_{asteroid}*V^2$) since the asteroid is vaporized on contact. Assuming the average velocity of a cosmic projectile in the Solar system is about 20 Km/sec, this blast has released $\sim1.5$ Megatons of energy (1 Megaton of TNT (Mt) = $4.184\times10^{15}$ Joule), or about the size of a large nuclear weapon explosion. This can then evolve to an introduction on climatic change, species mass extinction, etc.

Using Newton’s law ($F=ma$) and data from Table I, assuming planets are spherical, students can estimate the gravitational acceleration, $g_P=GM_P/R_P^2$, of these planets, their velocity through Kepler laws and its typical kinetic energy.
\begin{table}
\label{tab:sample}
\centering
\begin{tabular}{|l|l|l|l|l|}
\hline
 & Radius (km) & Mass (1024 kg) & Period rot (days) & g (m s-2)\\ \hline
Venus &6052 &4.8985 &243 &8.87\\
Earth &6371 &5.9736 &1 &9.81\\
Mars &3390 &0.64185 &1.027 &3.73\\
Moon &1737 &0.07349 &27.3217 &1.62\\
Io &1821.5 &0.0893 &1.77 &1.80\\
Europe &1561 &0.048 &3.55 &1.31\\
Ganymed &2631 &0.1482 &7.15 &1.43\\
Callisto &2410.5 &0.1076 &16.69 &1.24\\
Titan &2575 &0.1345 &15.95 &1.35\\
Encelade &252 &$1.08\times10^{-4}$ &1.37 &0.11\\
Phobos &22 &$1.072\times10^{-8}$ &0.319 &0.005\\
1999 KW4 &1.5 &$2.353\times 10^{-12}$& 0.1152 & 0.00043\\
Gaspra &12 &$10^{-8}$ &0,2955 &0.014\\
\hline
\end{tabular}
\caption{Physical characteristics of Planets, satellites and asteroids used in the exercises. }
\end{table}


The rim radius $R_c$ of a crater formed in a collision between an asteroid and a planet has been estimated to scale as $R_c= 10.14 G^{-0.17} a^{0.83} U^{0.34}$ m,  where $G=g_{asteroid}/g_{planet}$ is the ratio of the gravitational acceleration, $a$ is the asteroid radius in meters and $U$ its velocity in km/s.
With data and measurements, using the above equation, students derive some
characteristics of the impactors and compare to other cases in the pictures. Bottom images, the first image presents a view by Cassini of Mimas, a satellite from Saturn, during its January 2005 survey. Its diameter is 392 km. The huge crater present on the surface of this satellite has been named after Herschel who discovered it in 1789. The second image is a microscopic Moon spherule imaged with an electronic microscope. This spherule has been brought back by the Apollo 11 mission and its origin is a meteorite which has impacted the Moon: the energy of the impact melted some of the splattering rock and a fraction of it cooled into tiny glass beads. A microscopic crater is visible on the upper left, surrounded by a fragmented area caused by the shock waves of the small impact. The horizontal line in the bottom of the image scales as 100$\mu$m. Classroom may then consider the following steps:
\begin{itemize}
\item Describe both images and compare them. Discuss and determine the diameter of each crater. Give the ratio between the diameter of the crater and the diameter of the object (a rough scale of 20:1), satellites and spherule. Discuss what type of body has been able to produce such craters. Why their form is so different? Which type of crater is most frequent?

\item Open a Ganymede image from the space mission Galileo in 1997. Again consider the type of body and configuration that may have produce such features. Consider a scenario similar to the impact of Jupiter with the comet Shoemaker-Levy-9 in 1994. Estimate the energy of impact by calculating the asteroid kinetic energy ($E=\frac{1}{2}m_{asteroid}*V^2$), assuming the asteroid is vaporized on contact. Compare to Earth big craters and put things into perspective : the Chicxulub impact structure in Yucatan Peninsula, in Mexico, has a diameter of about 200Km, was produced 65 million years ago at the end of Cretaceous Period and is thought to be linked to the dinosaur extinction. 
\end{itemize}

\subsection{Kepler laws: The Galactic Black Hole}
The Milky Way galaxy is composed of approximately 100 billion stars and of numerous clouds of gas. Its shape is a disk – approximately 80,000 light-years in diameter – with a bulge nucleus in its center. In this exercise students will use the European Southern Observatory 8-meter Very Large Telescope (VLT) data and reproduce the calculation of the mass of the black hole in the Center of our Galaxy (Fig.4). This exercise has proven to be a blast among students as it triggers their imagination and creativity.  The exercise starts with a brief historical introduction and some explanations about the morphology of our Galaxy, to new distance concepts and black hole literature, both scientific and fictional. Students start to learn the concept of light-year (distance travelled by light during a year) and light-day, the distance travelled by light during a day, as most appropriate units to measure cosmic distances. Light-days are indeed the order of magnitude of the orbital periods of stars close to the galactic center. This Milky Way black hole region is also known for its intense compact radio source, Sagittarius A (SgrA*), one of the most bright radio sources in the sky. The application of Kepler Law’s to the motion of stars around the unseen object in the center of our galaxy, a massive object that is believed to be a black hole with a mass of 4 million solar masses,  astronomers were able to determine the mass of the supermassive black hole. A stack of images of the galactic centre region covering about 16 years of data enabled the follow-up of the motions of the most central stars. The plan may proceed as follows:
\begin{figure}
\centering
\includegraphics[width=5.5 in]{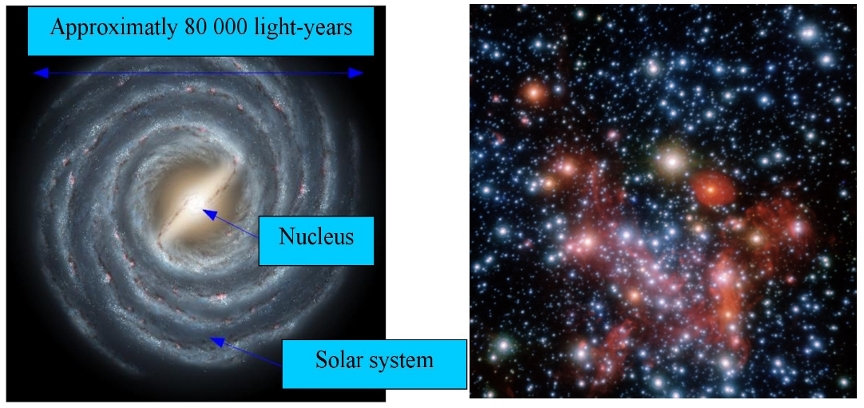}
\caption{Left: The Milky way structure; Credit: R. Hurt (SSC), JPL-Caltech, NASA. Right: The central parts of the Milky Way, as observed in the near-infrared with the NACO instrument on ESO's Very Large Telescope. @ESO.}

\end{figure}

The whole set of 16 years of VLT images to be analyzed is available at the site of the exercise for download. The first part will build an animation of the very same sky field with the image inbuilt software. This will produce an immediate evidence that those central stars are moving around some unseen object. Analyzing each star and measuring its motion from one image to the other it is possible to evaluate its orbital period, the time it takes to revolve around the black hole.
\begin{itemize}
\item {\bf 1st Kepler law} i) ``The orbit of a small body around a massive object is an ellipse with this object at one of the two foci." Students start selecting the S2 star (Fig.3). ii) Plot star pixel position as a function of time and fit an ellipsis to the points to determine semi-major and semi-minor ellipsis axis ($a,b$) and its foci ($c$) positions using the relation $c = (a^2-b^2)^{1/2}$. Scale pixel sizes to light-days and thus obtain orbit axis physical dimensions.  1st Kepler law enables the determination the position of the celestial body (SgrA*) that attracts the S2 star at one of the foci. Compare to a movie made after these SgrA* image stacking, where the elliptical movement is explicit. From the pixel positions plot it is possible to determine the orbit period $T$, in years.

\item {\bf 3rd Kepler law} Using the Kepler Law: $T^2/ a^3= 4 \pi^2/ GM$, where G is the gravitational constant $G$, $a$ is the elliptic orbit semi-major axis and $M$ is the mass of the central body. From here, it is easy to invert the equation and calculate the black Hole mass $M = 4\pi^2 a^3 / G T^2$. As a first approximation the plan of S2's orbit shall be taken as perpendicular to the observation line of sight. One can then compare M with the mass of the Sun ($M_{Sun} = 2\times 10^{30}$ kg) and learn on how to give the mass of these very massive object in terms of ``solar units''. As an example, the S2 star has a revolution period of 14 years, and the obtained results are close to the 4 million solar masses.

\item{\bf Further exploration: simple General Relativity} Students are encouraged to explore a scientific paper\cite{sgrA} on the Galactic Black Hole properties and answer topics like : i) If the plan of S2's orbit is not perpendicular to the line of sight estimate if the real mass of Sgr A*. Is it bigger or smaller than the one previously determined and how this may affect its properties? ii) Compare SgrA* mass to the one found by those researchers. For spherical black holes, the Schwarzschild Black Hole radius is the black hole maximum radius, delimiting the sphere from which nothing can escape. It is equal to $ R_S = 2GM/c^2$, M is the mass of the black hole. It is then possible to deduce the Sgr A* maximum radius. Compare values to the one scientifically obtained.

\end{itemize}
\begin{figure}
\centering
\includegraphics[width=7.0 in]{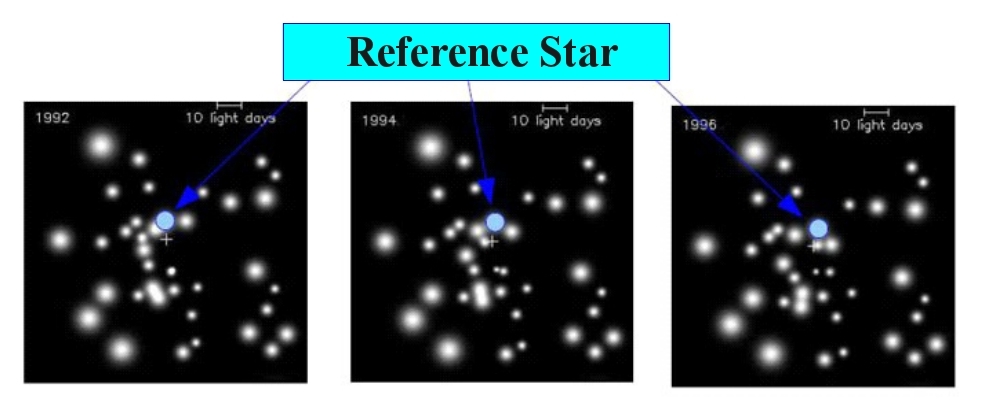}
\caption{The SagA* galactic field and the S2 reference star (@ESO). Here the images show 5 years of the 16 years of data images available and it is visible the change in position of S2 due its orbital movement around the galactic center.}
\label{fig3}
\end{figure}

\subsection{Hunting exoplanets}
Finding planets outside solar system - exoplanets - has become one of the most important science cases in astronomy. Exoplanet observations have opened new research avenues on planet formation and the possibilities of life in the Universe. Astronomers have made detections of 755 exoplanets by the end of January 2012\cite{exodatabase1,exodatabase2} and this number is continuously growing every week. The vast majority have been observed through indirect methods like radial velocity observations and transit methods rather than actual imaging. Most are giant planets resembling Jupiter with a wide variety leading to a ``taxonomic'' classification by mass, nature, size, etc. The huge distances between observer and the planet parent star make detections hard. Nevertheless, among the different methods of detection the most effective ones are:
\begin{itemize}
\item {\bf radial velocity:} the first optical method used to detect an exoplanet\cite{queloz} around a solar-type star. It is usually the detection choice from ground observatories and still the most efficient one. It provides information about the mass of the exoplanet through the use of Doppler spectroscopy and enables the use of Kepler laws for orbital parameters determination. On the EU-HOU website you can find a high-school level exercise about the detection of exoplanets by this method.
\item {\bf the transit method:} complementary to the radial velocity method. By studying the luminosity variation of the star when the planet passes in front of it, the radius of the planet can be determined; this is the method of choice used by the NASA Kepler mission\cite{kepler} and the French Corot satellite\cite{corot} to hunt exoplanets. 
\end{itemize}

\begin{figure}
\centering
\includegraphics[width=6.0 in]{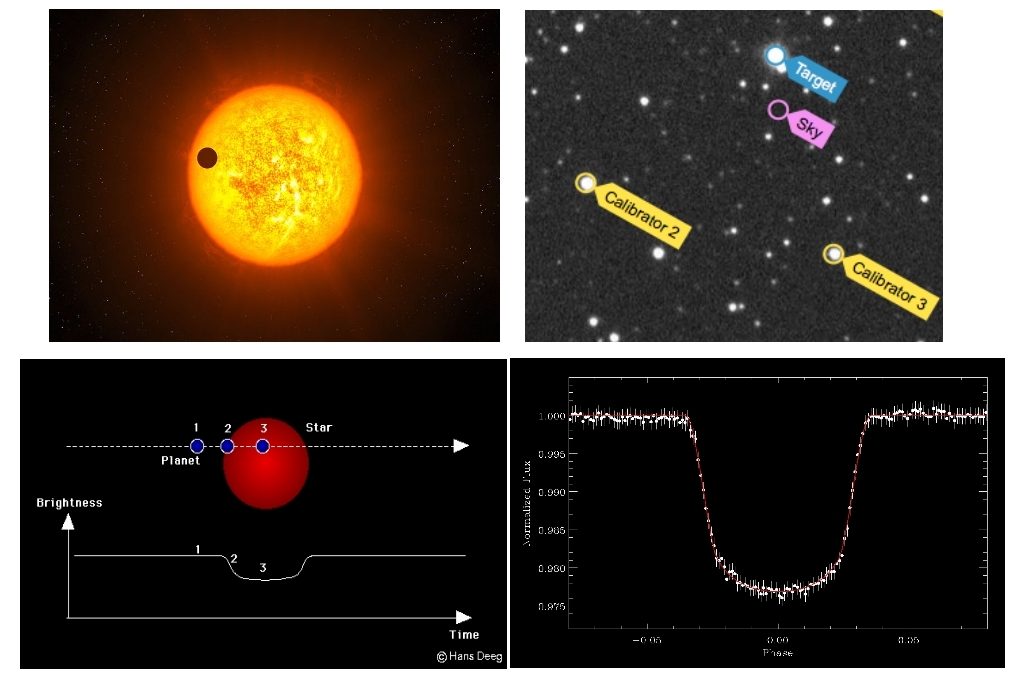}
\caption{From left to right, Top to bottom: Artist impression of planet passing in front of a star (CoRoT, ESA). Exoplanet hunting from a NASA Infrared Spitzer Space Telescope image through SalsaJ. Planet transit in front of star: as the planet passes in front of its parent star, the brightness of the star decreases (@ Hans Deeg, from 'Transits of extrasolar planets'.) The stellar light curve shown is the one of Corot-1 exoplanet star (Corot, ESA).}
\label{fig:exo}
\end{figure}
The prepared exercise starts with an introduction of these most important methods used by researchers to discover exoplanets. Students will use a set of 20 images taken by the NASA Spitzer Space Telescope\cite{spitzer}. Finally this activity is complemented by the tools available through SalsaJ to datamine data from space-base observatories. Actually, this ``Citizen Public Science'' has discovered two exoplanets in the Kepler Public Archive Data that passed unnoticed to scientists\cite{fischer}. Complementing this activity, the Las Cumbres Observatory Global Telescope Network\cite{lcogt} invites people to help understand better the physical characteristics of planets around their star with many ground-based datasets, through {\bf Agent Exoplanet}\cite{exo}. For schools these are unique user friendly opportunities to gracefully introduce their students to the world of research, engaging students in the measuring techniques of the star brightness changes with one or more orbiting planets.

Specifically, the mission starts with the choice of a planet-star system to be observed, through the following steps:
\begin{itemize}
\item A set of Spitzer infrared images enabling the study of exoplanet HD 189733b*\cite{hd18} will be distributed to the student who will then proceed to analyze the star suspected of possessing a planetary companion (as expected, we can not see the planet, but can only spot the change in its parent star brightness due to the occultation caused by the planet transit). This hot Jupiter extraplanet is at 63 light-years and was the first extrasolar planet to be mapped\cite{hd18a} and the first to be discovered with carbon dioxide in its atmosphere. These images are spanning several hours in order to cover a complete transit in front of the star(on Fig.5c), identifying the measured star flux with exoplanet position.

\item Students will choose three different stars and measure their light curves (or flux variation). They will proceed to check the non variable stars and properly eliminate all features not related to the transiting planet (Fig.5b). Only one of these stars (hosting an exoplanet) will show a variable light curve.

\item A full treatment of the star brightness dip will indicate the relative size and distance of the planet, compared to its parent star. The diameter of the parent star is known to be $0.788\pm 0.051$ solar radius. By carefully measuring the transit of the planet through the ``edge'' of star (points 1 and 2 at Fig.5c), estimate the diameter of the planet and its relative distance to the star using Kepler laws.

\item Agent Exoplanet\cite{exo} allows students to take the project further by providing many more datasets for transiting extra-solar planets, which can either be analyzed with SalsaJ or the Agent Exoplanet web-based photometry.
\end{itemize}

\begin{figure}
\centering
\includegraphics[width=6.6 in]{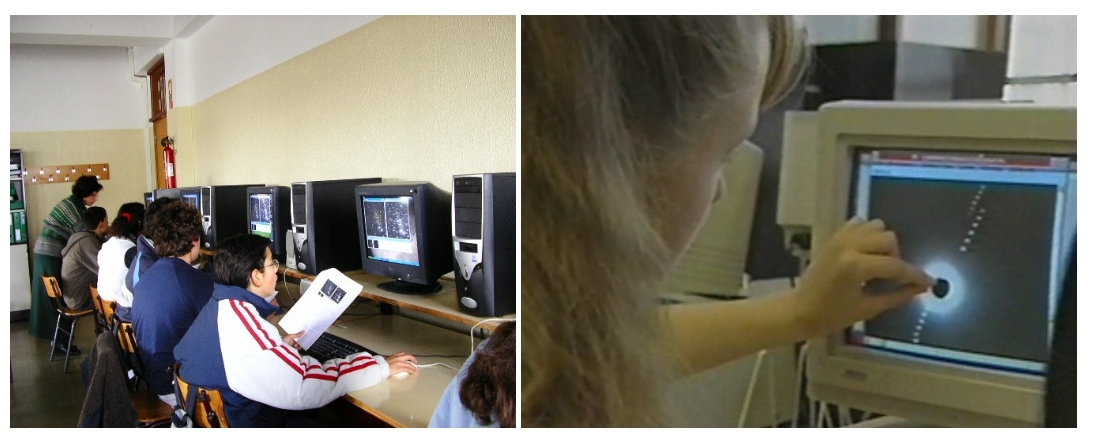}
\caption{Left: Hunting the supermassive Galactic Black Hole in Portugal. Right: determining Keplerian motion from Jupiter moons in a US classroom.}
\label{fig2}
\end{figure}
\section{Conclusion}  Research quality astronomy images are now abundantly available on the web. SalsaJ offers a new possibility: the free use of these amazing astronomical images from ground and space observatories for processing and physical interpretation. Every image taken by the most important astronomical observatories on Earth and Space covering most wavebands eventually end up in the open archives, freely available on the web. This is a genuine scientific treasure as they are the gateway to new science: science that can be done by the students in classroom revising the archives of professional astronomers in search for especial answers or by curious citizens trying their luck with projects called "citizen science projects". 

SalsaJ software has been developed as a powerful platform to support the teaching of sciences in secondary schools, and in particular dig the wealth of astronomical data connected to Virtual Observatories for the benefit of Physics teaching. In this paper we list some of the of the most popular exercises existing in the present plethora of tutorials available for SalsaJ exploitation. They cover many of the physics concepts taught in middle and high schools science curricula.

\begin{acknowledgments}
SalsaJ is developed and maintained by the University of Paris Pierre et Marie Curie. We are most grateful to many people working with the partners of the project and their collaborators, who support the SalsaJ development at various levels and refined through many iterations the exercises and tools using it. We acknowledge Fabrice Mottez, Weronika Sliwa and Rachel Comte by their contribution for Measuring distances with Cepheids exercise. It is a great pleasure to thanks Richard Bonnaire, who has initiated the SalsaJ project, David Weissenbach, who has started the modification of the interface and Reynald Berriot, who has helped to stabilize the code at an early stage and J\^erome Lucas, Olivier Marco. Lastly, J.L. has been hired thanks to the MINERVA grant to strengthen the existing structure, develop the spectra visualization tools and produced a manual. We also thank M. Metaxa and A. Bobetsi (Philekpadeftiki Etairia), I. Bellas-Velidis, A. Dapergolas (National Observatory, Athens) for developing the spectra and exercise plugins, Lech Mankiewicz (CFT, Poland) and A. Kalicki (Warsaw Technical University)for developing the webcam plugins. The development of SalsaJ has been initiated thanks to the support of the European Space Agency, the {\it LUNAP - Science \`a l‘Ecole} project of the French Ministry of Education. The ``EU-HOU: Hands-On Universe, Europe. Bringing frontline interactive astronomy to the classroom'' project EU-HOU has been initiated through a MINERVA grant from EC, then supported by a Comenius one.  The US Hands-On Universe has been supported by generous funding from the US National Science Foundation, through grants ESI-9252915, ESI-9554161, and ESI19819579. DB is supported  by a Ci\^encia2007 Research Contract, funded by FCT/MCTES (Portugal)and POPH/FSE (EC).
\end{acknowledgments}



\end{document}